# Lisp, Jazz, Aikido

## Three Expressions of a Single Essence

Didier Verna[a]

a   EPITA Research and Development Laboratory

**Abstract**   The relation between Science (what we can explain) and Art (what we can't) has long been acknowledged and while every science contains an artistic part, every art form also needs a bit of science. Among all scientific disciplines, programming holds a special place for two reasons. First, the artistic part is not only undeniable but also essential. Second, and much like in a purely artistic discipline, the act of programming is driven partly by the notion of aesthetics: the pleasure we have in creating beautiful things.

Even though the importance of aesthetics in the act of programming is now unquestioned, more could still be written on the subject. The field called "psychology of programming" focuses on the cognitive aspects of the activity, with the goal of improving the productivity of programmers. While many scientists have emphasized their concern for aesthetics and the impact it has on their activity, few computer scientists have actually written about their thought process while programming.

What makes us like or dislike such and such language or paradigm? Why do we shape our programs the way we do? By answering these questions from the angle of aesthetics, we may be able to shed some new light on the art of programming. Starting from the assumption that aesthetics is an inherently transversal dimension, it should be possible for every programmer to find the same aesthetic driving force in every creative activity they undertake, not just programming, and in doing so, get deeper insight on why and how they do things the way they do.

On the other hand, because our aesthetic sensitivities are so personal, all we can really do is relate our own experiences and share it with others, in the hope that it will inspire them to do the same. My personal life has been revolving around three major creative activities, of equal importance: programming in Lisp, playing Jazz music, and practicing Aikido. But why so many of them, why so different ones, and why these specifically?

By introspecting my personal aesthetic sensitivities, I eventually realized that my tastes in the scientific, artistic, and physical domains are all motivated by the same driving forces, hence unifying Lisp, Jazz, and Aikido as three expressions of a single essence, not so different after all. Lisp, Jazz, and Aikido are governed by a limited set of rules which remain simple and unobtrusive. Conforming to them is a pleasure. Because Lisp, Jazz, and Aikido are inherently introspective disciplines, they also invite you to transgress the rules in order to find your own. Breaking the rules is fun. Finally, if Lisp, Jazz, and Aikido unify so many paradigms, styles, or techniques, it is not by mere accumulation but because they live at the meta-level and let you reinvent them. Working at the meta-level is an enlightening experience.

Understand your aesthetic sensitivities and you may gain considerable insight on your own psychology of programming. Mine is perhaps common to most lispers. Perhaps also common to other programming communities, but that, is for the reader to decide…

# The Art, Science, and Engineering of Programming

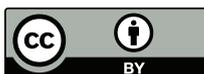

## 1 Prologue

### 1.1 Strange Attractors

What could computer science, music, and martial arts possibly have in common? For as long as I can remember, they have been like "strange attractors" to me, like three pillars around which my whole life evolves and revolves.

No matter how much distance I may have put between them and me in the past, I always ended up coming back to them, every time with a feeling of relief after the sudden realization that I was losing my way. Not unlike that weird sensation of gently falling asleep and then being hit by a brutal sparkle of consciousness. Or awakening from an anesthesia without even remembering having passed out.

I can also remember very precisely the excitement I felt when I first encountered the Lisp family of programming languages, when I was first introduced to Jazz music and improvisation, and when I had my first practice session of Aikido* (a Japanese martial art[1]). Different things, same emotion. And also the feeling that in some way, I was meant to be a Lisper, a Jazzman, and an Aikidoka* (an Aikido practitioner) before knowing these disciplines even existed.

But why?

### 1.2 Genesis

Some time ago, I had the immense pleasure of reconnecting with one of my primary school teachers, after several decades (I was around 10 at the time). I think every one of us have this one very special teacher, who marked us for life, and whom we hold very dearly in our hearts. It is from those two years of primary school I spent with him that I retain the most wonderful memories from my childhood.

In particular, and this is no coincidence, this teacher had a way of attaching as much importance to creative activities in the class, as to more regular and mandatory trainings at this level, such as reading, writing, and counting. If you finished an assignment early, you had the right to go to a special shelf and pick a puzzle or a book of riddles to play with, waiting for the others. He also had established a daily slot, that he called "The Interview", in which every one of us could do absolutely anything we wanted, and present it to the rest of the class.

I remember using a fair share of this slot, doing sketches, jokes, or miming my favorite musicians at the time, with a plywood guitar I had tinkered. Even though I believe that creative or artistic skills, like every other talent, are innate more than acquired, I also do believe that I owe this teacher a lot in that regard, be it only for having provided me with a very early channel through which I could express my creative urges, let them blossom in a watchful and benevolent environment, and exercise that utter freedom of expression that every artist is so profoundly attached to.

---

[1] *Note: terms followed by an asterisk are Japanese and are listed in the glossary in appendix A.*

While we were reconnecting with each other, the conversation naturally moved on to our personal lives, and I started to describe my regular week, split in two. Half of the time being a computer scientist, the other half of the time being a musician; I indeed could never choose between the two careers, so I ended up doing them both.

To which he looked at me with a teasing face, and said "So you've essentially become totally schizophrenic!". This was of course a friendly and innocent joke misusing, as often, the word "schizophrenia". What he really meant was to ask, although not seriously, whether I was suffering from MPD (Multiple Personality Disorder), *a.k.a.* DID (Dissociative Identity Disorder), or even just *split personality* for short [86].

But as we laughed at the joke together, a curious thought started to grow in my mind: what if there was some amount of seriousness in the joke? Why could I never choose one path over the other(s)? Could I actually be suffering from some kind of mental illness that would make me live two, or even three different lives? If not, if I'm only a single coherent self, then there had to be some kind of unification behind those three strange attractors of mine.

That is basically how the reflection that lead to this essay started. Mostly as a work of introspection, slightly tainted with a drop of anxiety. What could computer science (Lisp), music (Jazz), and martial arts (Aikido) possibly have in common, that would confirm my mental sanity?

## 2  Introduction

Science: an Oedipal son of Art. It could only liberate itself by perverting its parent. As related by Knuth [57], Art was originally the mother of all knowledge, meaning "skill" in Latin, with its equivalent Greek meaning "technology". If there were to be a distinction in these early days of civilization, it would have been between *liberal arts* (the arts of the mind, including things we would now label as Science like logic) and *manual arts*. When Science emerged, the meaning of Art got perverted: Science came to mean knowledge, and Art application of knowledge (the *craft* of the *art*-isans). Knuth also pushes the distinction between Art and Science a bit further: Science means knowledge that is completely understood, and Art, something not fully understood {1}.[2]

### 2.1  Art and Science

Because we are very far from a full understanding of the world we live in, it is commonly accepted now that every discipline needs both Science and Art, something also pointed out by Knuth. For example, even in Mathematics, perhaps the purest form of Science, we do not fully understand the process by which elegant solutions to difficult problems may be "discovered", or why some proofs appear more beautiful than others, so there is a non-negligible part of Art in the Science [43]. Conversely, there is a proportion of Science even in fine arts. Although we don't really know why we are more touched by some music or artist than by some other, there are theories of harmony, rhythm, even improvisation, on which musical discourses are constructed. The same applies to martial arts. Aikido, for instance, is based practically on a single, somewhat mysterious and magical fact: relaxation beats strength. Nevertheless, for everyday practice, it helps to know the importance of your center of gravity (Seika Tanden*) and its position in space, it helps to move your feet using one of the codified ways (Tai Sabaki*) *etc.* In other words, the way to the mysterious heart of the discipline is paved with concrete techniques, nomenclature, and codification.

---

[2] *Note: numbers in curly braces (e.g. {42}) refer to quotations listed right after the bibliography.*

## 2.2 Programming

Among all disciplines mixing Art and Science, programming holds a special place, and at the risk of going against a general trend of the time [6], Knuth was not the only one to think that the artistic aspect of programming is not only desirable, but also essential. Two other notable advocates of the "art of programming" are Dijkstra, who tried to demonstrate that Art in programming, far from being a luxury, usually pays off [29], and Ershov, who went as far as claiming that there is an inherent system of values in programming that relates more to Art than to Science [34]. But perhaps the most striking point in Knuth, Dijkstra, and Ershov's arguments is that the artistic aspect of programming deals with such notions as elegance, aesthetics, beauty, style, pleasure, and emotion {2,3,4}.

## 2.3 Aesthetics, Beauty, Pleasure

We won't risk playing the philosopher by defining beauty and aesthetics here. Rather, we simply choose to stick with the definitions given by Morris [70]: beauty is that which gives pleasure in use and contemplation, aesthetics is the discipline studying it or the process to achieve it. These definitions seem to be in total conformance with their use by Knuth, Dijkstra, and Ershov.

My original anxiety for a potential MPD affection was slightly relieved by learning about Knuth, Dijkstra, and Ershov's care for the aesthetics of programming. Just like playing music, and without any consideration for usefulness {5}, the act of programming would be driven, at least partly, by the pleasure of creating beautiful things {6,7} (what about martial arts though?). But we can continue to dig deeper and explore the aesthetic dimensions involved. Perhaps with the notable exception of Hofstadter [49] and, to a lesser extent, Graham [41], writing about the transversality of aesthetics is rare in the literature, especially in relation with computer science...

> *the transversality of a concept or a thing is the characteristic of having aspects that cut across other concepts or things — cross-cutting aspects; aspects that sit athwart {8}*

...In fact, Hofstadter's book is indeed transversal, but about the emergence of cognition rather than about aesthetics. The field called "psychology of programming" focuses on the cognitive aspects of the activity, with the explicit and concrete goal of improving the performance and productivity of the programmers [48, 106]. While many scientists have emphasized their concern for aesthetics {9}, few *computer* scientists have actually written about their thought process while programming {10}.

I can see three reasons for this. First, it may be frightening to do so. Reflecting on our aesthetic sensitivities involves thinking about our own emotions, and not everyone is ready or willing to perform that kind of introspection. Second, this introspection needs to be very deep, because our aesthetic sensitivities are extremely personal. Third, this introspection also needs to be very broad, because our aesthetic sensitivities affect every part of our lives.

So yes, the relationship between the act of programming and Art is very strong (it is with martial arts as well; this will become clearer soon), but which relationship is it exactly, and on a personal note, why does it lead to Lisp, Jazz, and Aikido specifically? Lisp, Jazz, and Aikido, as it turns out, are three very different disciplines, yet ruled by the same set of aesthetic dimensions, hence complementary. To paraphrase Hofstadter [49]: "I realized that to me, Lisp, Jazz, and Aikido were only shadows cast in different directions by some central solid essence. I tried to reconstruct the central object, and came up with this essay."

## 3 Rules, Games, Players

Systems and rules are central to our lives. As matter, we abide by the laws of physics. As living beings, we abide by the rules of biology. As humans, we are educated to certain moral values and we learn how to behave in society. Games and play are equally central…

> *a game is a mental or physical activity based on a system of rules, the purpose of which is nothing more than the pleasure it procures*

…Etymologically speaking (Latin), playing is connected to amusement (*jocus*) but also learning (*ludus*) and it's very likely that playful behavior plays a major role in the evolution of life. Indeed, since the classic 1929 experiment of Blodgett [11], we know that playing is an exploratory behavior leading to either an increase of knowledge, or an acquisition of skills useful for survival. That is why playing is crucial, in particular for offspring development, and by that, I don't mean only for humans.

So it appears that we are all players and that this behavior is hard-wired, to some extent. But what are the rules?

### 3.1 Programming

In programming, the very first set of rules is of course the syntax and semantics of the language. The computer as a whole is an incredibly large set of rules in which the programmer has to evolve, and, as noted by Ershov [34], programming suffers from the particularity that 100% accuracy is required, contrary to other professions that may use different phenomena (e.g. biological or physical) without fully understanding or mastering them. In many ways, a programmer *playing* with a new programming language is just a kid trying a new game. Programmers *do* use the verb "to play" in such circumstances, which is no coincidence.

## 3.2 Music

Just as programming comes with a great diversity of languages (hence rules and systems), music of course comes with a great variety and number of rules and systems as well. Occidental music is based on the so-called *diatonic* (heptatonic) scale, most of the time tempered, but some other music — called *micro-tonal* — will use more, possibly unequal intervals. This is the case for many traditional music forms (Celtic, Arabic, Indian, Oriental *etc.*). As far as rhythm is concerned, the vast majority of occidental music can be expressed in 2/4 (marches), 3/4 (waltz), and 4/4. Some other music, however, will use so-called *asymmetric measures* (e.g. 7/4 in central Europe and Russia). The presence of asymmetric measures in occidental music is exceptional (progressive rock, and a few classical composers such as Erik Satie and Béla Bartók).

At a higher level, there are rules for composition too. Classical music, in particular, obeys very strict rules (you can't call just anything a symphony or a sonata). Even more modern popular music follows common compositional patterns. A lot of Jazz songs are based on the so-called *AABA* structure, a structure repeated as many times as needed to first expose the melody, then leave room for choruses, and at the end expose the melody again, one last time.

## 3.3 Martial Arts

While in sports every game has its own set of rules, true martial arts normally don't. That is because both theoretically and historically, they deal with life and death situations, so anything is possible. This point is crucial to understand the difference between combat sports and martial arts. Judo*, for instance, is not a martial art anymore. By introducing competition in Judo, the art got distorted and became a combat sport, with weight categories, specific timing, and other rules that simply don't make sense in an actual conflict situation…

> *sorry, you can't molest me: we do not belong to the same weight category*

…So if anything is possible in a traditional martial art such as Aikido, what do we conform to? The key is that even if anything is possible in theory, you still need to train on particular forms: the Kata*. In Aikido, just like in any other martial art, there is a more or less fixed but at least properly documented set of techniques for attacking, responding to an attack, moving, projecting or immobilizing one's partner, applying joint pressure *etc.*

In order to link this to programming or music, consider that a Turing-complete programming language lets you express any algorithm you want, although the language itself is limited by its grammar and semantics. Similarly, the number of composable symphonies is endless, although the musical language in use is very constrained. Martial techniques are equivalent to a programming language syntax and semantics or a musical style: they do not limit your expression — perhaps only make some things more difficult to express than others. On the other hand, the limitations imposed by competition in Judo is more equivalent to a programmer being forbidden to write some particular programs, or a composer not having the right to write a specific symphony.

Finally, at a higher level, every martial art also provides different layers of rules explaining what is expected from the practitioners. In the Japanese martial arts (Budo*), and particularly in Aikido, the Reishiki* is a code of conduct describing the ceremonial aspects and other concrete behavior that should be obeyed. On top of that, the Reigisaho* (the etiquette) emphasizes the moral values that the Reishiki helps develop, and that apply not only in the Dojo*, where we practice, but in life in general. There, we find such values as respect and courtesy. Although the Reigisaho is traditionally maintained by oral transmission, some of our masters have written about it [83].

*Which postures can we adopt in front of systems and rules? Which forms of pleasure can we get from the adopted behaviors, identical in Lisp, Jazz, and Aikido?*

## 4 Aesthetics of Conformance

The first behavior we can adopt in front of systems and rules is obviously to honor them. There is pleasure in evolving within systems and rules, but more than that, there is also pleasure in making the most out of the limitations or constraints we face. This is what we call *aesthetics of conformance*, probably quite universal a dimension.

### 4.1 Playing by the Rules

In programming, the aesthetics of conformance (and the beauty that it leads to) lies in our ability, as programmers, to adapt our ideas to the limitations of the language we're using. A particular notion, the *programming pearl*, probably describes the paroxysmal level of pleasure that we get from the aesthetics of conformance in programming. Programming pearls are so important today that many conferences organize tracks for them and people write books about them [8]. Programming pearls are "beyond solid engineering, in the realm of insight and creativity". In other words, they make the most of the tools available, and there is beauty in that.

Rules in music are obviously not as strict as those in programming. Musicians and philosophers alike consider that to be "functional", a good composition must make the listener oscillate between comfort and surprise [63]. The aesthetics of conformance, particularly important in classical music, exercises the element of comfort: the pleasure of hearing something expected. The beauty of it hence comes from your ability, as a musician, to adapt your ideas to the constraints of the music. As a composer, you must obey compositional rules specific to the music style. As a performer of written music, you must obey the score and make the most out of (your interpretation of) it. As an improviser, you must obey a style, pitch, tempo, rhythmic pattern, chord progression *etc.* Just like a programming language, a musical style, or even a specific score limits your expressiveness, and there is beauty in making the most out of these limitations.

Given what we said about rules in martial arts earlier, the aesthetics of conformance would consist in executing the techniques (the Kata) to perfection. One particular writing of the Japanese term Kata confirms this: 型. Written like this, the sense is "original and ideal form". In Aikido, the lower grades (Kyu*) deal precisely with the aesthetics of conformance. The jury evaluates the truthfulness of your attacks and the quality of your reactions, all of this from a purely technical point of view. Only later will you be evaluated on more profound criteria. Some martial arts have deviated from the original philosophy, to focus on the technique only. An example coming from Karate* is the so-called *Karate-Kata*, where the practitioner executes a series of codified movements without an actual opponent in front of him. Karate-Kata hence concentrates on the aesthetics of conformance exclusively. Hardly a martial art anymore, it can still be beautiful to watch. One of the reasons that made me fall in love with Aikido was the incredible beauty of it. From the outside or as a beginner, you don't have the ability to understand what's really going on, so you can only be sensitive to this first aesthetic dimension. By just looking at it, I find that Aikido is by far the most beautiful martial art there is; almost a dance, supple, fluid, obvious.

## 4.2 The Nature of Constraints

Conformance is probably a quite universal aesthetic dimension. Some aspects of it, however, drove me specifically to Lisp, Jazz, and Aikido. To understand why and how, it is worth investigating the different kinds of challenges that conformance offers.

It seems that the pleasure of conformance increases with the constraints: the more constraints, the greater the challenge. But what is the nature of constraints? This question doesn't seem to have been much developed, as even Knuth, who thought a lot about aesthetics in programming, doesn't have a clear view of it. For example, in his 1974 ACM communication [57], he says "One rather curious thing I've noticed about aesthetic satisfaction is that our pleasure is significantly enhanced when we accomplish something with limited tools." (*i.e.* the more constraints, the greater the challenge). However, later on he says: "Please, give us tools that are a pleasure to use [...], instead of providing something we have to fight with", which seems contradictory: if you don't have to fight, where is the challenge?

### 4.2.1 Adequacy
One way of feeling constrained by a system of rules is when you find them inadequate, ill-formed, or even broken — plain and simple. When asked about the beauty of working as a programmer for a bank, Knuth reported having had a hard time finding an answer, but eventually claimed [59, p. 92]: "In general, I believe there is a way to find beauty even in COBOL programming".

While in the Knuth arena, let us of course mention (La)TeX, a wholly inadequate and particularly ugly artifact when it comes to general purpose programming {11}. Yet, a lot of people (myself included) find pleasure in making something out of it.

Bruno le Floch for instance, a notable TEX programmer and implementer, among other things, of a LATEX regular expression engine, no less, describes himself as "hooked by the strange style of programming that TEX macros provide". Finally, and more recently, Gary Bernhardt provided a striking demonstration of the pleasure one can get out of JavaScript or Ruby's ultimate brokenness [9].

Being in the academy and hence having the luxury of choosing what I want to work on and with, and perhaps also having grown older, I think I lost interest in playing with inadequacy or brokenness over the years. I remember having found pleasure in writing shell or C code long ago. I remember vividly my last C program. I remember very well both the anger I felt having to deal with memory management and potential buffer overflows manually, but at the same time the pleasure I had in finding elegant solutions to these limitations. Just not anymore…

> TEX still resists (but for how long?) my progressive disinterest for playing with inadequacy. The reason probably is that it is still the best solution there is to get nice typography, which I am particularly sensitive to. I must confess however that as my TEX activities gradually shift from development to maintenance and repair [98], I am more and more thinking of rewriting a new typesetting system in a decent general purpose programming language [104, 105]

…On the other hand, one reason explaining my affection for Lisp (Common Lisp [3] in particular) is that I find it adequate: among all the languages I encountered, Common Lisp is the one that gets the least in my way. In that, I'm in agreement with Knuth: I prefer a tool that is a pleasure to work with; that I don't have to fight with.

It is hard to speak of (in)adequacy in music or Art in general, because the sole purpose (usefulness) of it is the pleasure it procures. Any piece of music can be considered adequate as soon as there is at least one person to enjoy it. The question of adequacy in martial arts is equally nonsensical — although a frequently asked question. Is such or such martial art adequate (efficient) in a real fight? You trained bare hands so what do you do in front of a knife? You train with a knife, but then what do you do in front of a gun, or an atomic bomb? Back in the middle ages, martial techniques could be considered adequate or not in such or such situation, but today, the point is different. In the modern sense, the purpose is education and self-development *through* martial techniques. This is what led Japan in the transition from martial *techniques* (Bu Jutsu*) to martial *arts* (Budo) during the Meiji* era. So just as a piece of music is adequate as soon as there is one person to enjoy it, a Budo is adequate as soon as there is one person to enjoy it for personal development.

### 4.2.2 Support and Collaboration

Adequacy thus means not getting in your way, or not having to fight against. Pushed even further, this concept leads to the notions of *support* and *collaboration*: actually going *your* way, or fighting *alongside* with (see also appendix B.1). From this new angle, the notion of adequacy regains its relevance to music and martial arts, and particularly to Jazz and Aikido.

Earlier, I mentioned the adequacy of Common Lisp to my needs, but more than that, I cannot keep track of the times the language has actually helped me. After more than ten years with it, there are still frequent occasions in which, facing a specific problem, I find that the standard (in spite of its old age [3]), already provides a solution. Not only does the language feel adequate, it also feels supportive. In some sense, the GC (Garbage Collector) and the debugger, embedded in every Common Lisp application, contribute to this sensation as well. The GC supports you by taking care of memory leaks. The embedded debugger makes the application more forgiving about your mistakes, by offering you a chance to recover rather than just crashing.

This feeling of support and collaboration turns out to be quite strong in Jazz. Music, in general, can be very unsupportive of you. Classical music for instance will not tolerate mistakes, and even if at some point, you feel like taking some liberty with the score, you can't. The rules get in the way and you would have to fight with them. Consider for instance the impact of a wrong note from the soloist in the middle of a concerto, or a missed snare drum hit in Ravel's Bolero… In Jazz on the other hand, and more specifically in improvisation, the whole point of the music is to be tolerant, or even better, supportive. This may sound paradoxical because Jazz is also one of the most evolved and rich musical forms, requiring a lot of virtuosity from the musicians. However, precisely because so much in Jazz improvisation is happening in real time, it's only natural to tolerate mistakes {12}. The supportive and collaborative nature of Jazz was brilliantly illustrated quite recently by Herbie Hancock (one of the brightest contemporary Jazz pianists) in a masterclass teaser {13} .

The same feeling of support and collaboration strongly emanates from Aikido as well, perhaps more than from any other martial art. The genius of the founder, Morihei Ueshiba, was to use martial — hence potentially lethal — techniques to develop an "art of peace", as he was calling it himself. For example, while in Karate you confront an *opponent* and learn to inflict pain, or cause damage (albeit only in theory), in Aikido, you practice with *partners* and learn to control them in a harmless way, preserving their corporal integrity. As a consequence, all the former martial techniques incorporated into Aikido were transformed into protective ones, and you never actually fight {14}.

The non-competitive aspect, the smooth and fluid look, and the idea of not fighting leads to a very frequent misinterpretation. People think that Aikido is complaisant and that the demonstrations are somehow arranged in advance. On the contrary, it is precisely because you practice with a partner that Aikido is very real. For instance, because Aikido works on the circulation of the Ki (the energy), an Aikidoka simply cannot do anything if the attack is not for real. This means that in order for Aikido to work (and demonstrations to be impressive), collaboration requires truthfulness instead of leniency. The only problem is that all of this is not *visible* from the outside in Aikido, because no harm is ever done to an opponent; not even in theory. You need to experience it in order to understand it.

### 4.2.3 Simplicity

In addition to inadequacy, complexity is another kind of conformance challenge. There may be pleasure in making the most out of a set of intricate or numerous rules. C++ or Perl, for instance, are complex languages, notably on the syntactic level. Yet, some people enjoy using them. Lisp, on the other hand, is simple. Consider that it only takes a minute to learn the two fundamental syntactic concepts of the core language: *atoms* such as `this-symbol`, `"that string"` or the number `2.5`, and *S-Expressions* denoting both lists of elements and function calls in prefix notation, such as (`func arg1 (list of things) arg3 ...`). That's about it, and it is precisely the simplicity of this design that fascinated me during my first encounter with a Lisp dialect (see also appendix B.2).

My taste in music (specifically as a composer) is in full coherence with this. The rules of composition can be quite complicated and numerous. In classical music, counterpoint and fugue are typical examples. In Jazz music, composition can also be complicated. The music for big bands is probably where you find the utmost complication. Looking at my own music in retrospect however, I realize that I have never been much into writing complicated things. The longest score I have may be 4 pages long only. It is for a song entitled Corridor, from my first album [96], and the complexity of it comes from the fact that the theme is played with very specific chords on the guitar, which I had to write down precisely, and also because I wanted this theme to be backed up with a specific double-bass part, which I also had to write down precisely (a sample page is provided in appendix D). Apart from that, the majority of my other compositions fit on one or two sheets of paper, with the melodic line (sometimes not even rendered in detail), the underlying harmonic progression, and a rough indication of the style (blues, swing, Latin, Bossa-Nova *etc.*). The simplest one I have, from my second album [103], is exactly 4.75 lines long (see appendix E). It is also worth noting that the several hundred so-called *Jazz standards* (the most essential classics of the Jazz repertoire) are written down in this very simple way in the various *Real Books* available [87].

Unsurprisingly, this general taste for simplicity also drove me toward Aikido. First of all, Aikido is one of the last traditional martial arts to remain non-competitive. This means that when we practice, we are free from arbitrary, artificial, and potentially complicated rules imposed by competition. Contrary to other martial arts such as the Nin Jutsu* which originally counted no less than 18 different disciplines, Aikido also voluntarily limits the number of techniques or tools in use. For example, we are not interested in mastering a lot of weapons. We barely use three: knife, saber, and spear, and even when we do, the purpose is not to master those weapons *per se* but rather, to refine bare hands practice. Granted, Aikido still has a fair amount of codified techniques. Our mother house, the Tokyo Aikikai, has edited several books describing dozens of attacks, responses, projections, immobilizations, and the combinations of those [91, 94]. Most local federations in the world also have their own pedagogical resources (books, videos *etc.*). However, as you progress in the discipline, you start to realize that the techniques themselves are not so important, as they all boil down to a single principle, evaluated when you pass the higher grades (the black belt's Dan*): the circulation of the energy (the "ki" in Aikido). That is why Aikido is in fact very simple. In some sense, the only rule there is is to let the energy flow naturally. Everything else is a technical detail aimed at working on this principle. This, again, is what constitutes the difference between a *Jutsu* (focused on the techniques themselves) and a *Do* (using techniques as a means to work on something more profound).

### 4.2.4 Easiness

There is a frequent confusion between simplicity and easiness (or between complexity and difficulty, which is the same thing). Many non-Lispers believe in the urban legend that programming in Lisp is complicated, whereas in fact, it's not. For simple things, Lisp is probably even simpler to use than most other general purpose programming languages. The real problem is that most people believing this are already biased toward C-like languages (so they think differently), and there is evidence showing that students introduced to Scheme as their first programming language, as it used to be the case at MIT and UPMC (Université Pierre et Marie Curie) for a long time, do not have any problems with it [17]. On the other hand, very early in history, Lisp made available and adopted peculiar (non-mainstream) computing models such as the mixing of programming / compilation / execution phases, or meta-programming through homoiconicity, which may be difficult to grasp coming from mainstream languages.

The same considerations apply to Jazz and Aikido. While the rules for playing Jazz can be very simple and few, improvisation is a notoriously difficult art which requires a lot of practice to master. In Aikido, even the simplest techniques are difficult to master. The greatest difficulty is to be able to handle a conflict situation without emotional heat and in total relaxation. This is difficult because in order to achieve this, we must "unlearn" the bad reflexes that our reptilian brain, along with our educational biases want to trigger. In Aikido, it is not uncommon to require 10 years of practice before reaching the black belt (first Dan).

In a not-so-paradoxical way, a fascinating aspect of Lisp, Jazz, and Aikido is that some of the difficulty in their exercise comes precisely from their simplicity…

*Conformance leads to beauty. But do we really want to remain conformant all the time? Perhaps a life of conformance would end up being too boring…*

## 5 Aesthetics of Transgression

What can we do with a system of rules, apart from conforming to them? The obvious answer is *not* conforming to them. Rules can be transgressed as much as they can be obeyed, but more importantly, they can be transgressed *willingly*. There is fun in violating the rules (see also appendix B.3). This is what we call the *aesthetics of transgression*.

### 5.1 The Roots of Transgression

Nelly Lacince [60] notes that transgression can be seen from two angles. The first one focuses exclusively on the violation of established rules, in the negative sense, and with the potential retaliatory measures that may follow (value judgments, reprimands, punishments *etc.*). The second one, more in line with Bataille and Foucault [5, 36], prefers to view transgression in a more neutral, or even positive sense. This is the view we prefer to adopt. Transgression is then seen as a way to go forward, beyond the limits, without denying them. More than that, the very act of transgression even reasserts the rules, because without rules, there is nothing to transgress.

### 5.1.1 Transgression in Education

A first positive aspect of transgression lies in education. Seen as an oscillatory movement going from within to beyond the limits, and back, transgression is a means to get a deeper understanding of the rules, precisely by exercising the limits they entail. In that sense, some level of transgression needs to be legitimized.

Finding the right balance between conformance and transgression is a delicate endeavor, something that we, parents or teachers, know all too well. From early childhood to teenage, we know that our children construct their own personalities by going through several transgressive phases. Finding the adequate response in front of such or such transgression is crucial to their development. Too little room for transgression and you get a docile, servile population [69]. In education, this leads to a normalization of minds in which the actual needs of the students are never fulfilled: those of the misfits are ignored because their transgressive behavior is reprimanded, and those of the fittest are also ignored because they are invisible [37]. Too much room for transgression (in other words, too permissive an education) and you get a population of disturbed, temperamental citizens who will have a hard time living in any society.

### 5.1.2 Transgression in Art, Cultures and Societies

Transgression also appears to be crucial in the birth and development of an art form and the cultures it leads to. The case of violence illustrates this very clearly. As explained by Bataille [5], while violence is normally proscribed in modern civilizations, Art, on the contrary invites us to embrace it (hence providing an outlet for it). Some authors even consider this to be a key to our evolution as a species [31]. By depicting violence and death in the Lascaux paintings for example, the forbidden becomes symbolized instead of exercised, turning it into it a matter of discussion, external to the artist's condition, and consequently a way to gain knowledge about oneself. It follows that this very early step in self-awareness may constitute the precise point at which we escape from our condition of mere animals, to reach that of human beings.

### 5.2 Lisp Transgressions

According to Graham, transgression is crucial for progress in all scientific activity, and is motivated by mere curiosity, the will of taking nothing for granted, and simply the benefits of the brain exercise {15,16}. Very early in Lisp history, a rather transgressive event occurred: the decision to let the system *purposely* leak memory. Indeed, the GC was only invented and added to it later on. As related by Gabriel [39, p. 7] however, this design choice was made out of a concern for not "complicating the elegant definition of [the symbolic] differentiation [algorithm]" [65]. Hence, we see again the importance of elegance and simplicity, and how transgression can be born from aesthetic concerns.

### 5.2.1 Extensibility

In terms of programming, and given the absolute determinism of the computer, it may seem odd at first to even just consider the idea of transgression. Earlier, we said that in programming, the aesthetics of conformance lies in our ability to adapt our ideas to the constraints of the language we're using. It thus follows that the aesthetics of transgression would be to adapt the language to our ideas instead, and Lisp goes a long way in that direction. One key to understanding this is the notion of extensibility [25] [95] [89, section 2] [32, section 8] [101, section 2], indeed, an important characteristic of Lisp. In the particular case of Common Lisp, extensibility can take on at least three important forms. First, the programmer can access the *reader* and modify its behavior, making syntactic modifications to the language possible. Second, the macro layer provides a meta-programming facility with control over parse trees and evaluation semantics. The combination of syntactic extensions and macros makes Lisp particularly well suited for writing embedded homogeneous DSL's (Domain-Specific Languages) [101]. Third, the meta-circular implementation of CLOS (the Common Lisp Object System) via the MOP (Meta-Object Protocol) makes the very semantics of the object layer extensible or even redefinable. The reader interested in technical examples is invited to read appendix C.1.

Of course, the Lisp transgressions are possible only because the language not only allows them but supports them by providing us with transgression tools. In that sense, there are rules to transgress the rules, and with some experience, adapting the language to your needs becomes an integral part of the art of programming; the transgression becomes a rule in itself. Here we see again the positive aspect of transgression: a way to exercise the limits and push them away.

### 5.2.2 Introspection

I have always been fascinated by reflection and meta-levels in programming, probably because these notions are deeply related to that of self-awareness, and I am very fond of introspection at a personal, psychological level. It seems to me that transgression in programming is but one path on the quest for oneself. Very early in the history of programming, scientists underlined the importance of finding one's own style {3,17}. In most languages, the freedom of style is strongly constrained by the rules of the language. In Lisp however, it becomes apparent that we have more liberty to find not only our own style, but our own language within the language {18,19}.

Lisp becoming a different language for every programmer (sometimes even for every application) does not go without serious drawbacks, community-wise or in terms of language popularity (the community of lispers is indeed much smaller than that of C++ or Java). In addition to the explanations offered by Tarver [84], there is another fundamental reason. When we develop our own Lisp, we are essentially on a highly egocentric endeavor: the quest for a language perfect for us and only for us, which may be found only if we gain a deep understanding of our true self, completely disregarding the outside world. Programming Lisp, and in Lisp, is thus a highly introspective journey.

### 5.3 Jazz Transgressions

A good piece of music makes the listener oscillate between comfort and surprise, and the aesthetics of conformance exercises the element of comfort. It thus follows quite naturally that the aesthetics of transgression exercises the element of surprise…

> *surprise is a reaction to the unexpected*

…Whatever the style of music, a composer may decide to break a rule in order to produce an unexpected musical event that will surprise the audience. One may for instance suddenly change the rhythmic pattern, the key, the modulation (e.g. suddenly switching from major to minor) *etc.*

Many musical genres (classical included) use some level of transgression. Jazz, however, is unique in this matter, because the notion of transgression occupies a central place. I would even dare say that the whole purpose of Jazz is to transgress. Consider *Free Jazz,* for instance, a trend aiming at transgressing even the most liberal rules of Jazz at the time (the 50's), and in which saxophonists went as far as exploring alternative ways to blow in their instruments…

#### 5.3.1 Extensibility

To understand why Jazz is inherently transgressive, we can look at it from two different angles: its compositional and improvisational aspects.

From the compositional angle, the history of Jazz tells us that it cannot be defined as, or reduced to, a particular style of music. Listen to some Rockabilly, Twist, or Minuet. Pretty much the same today as on day one. Musical genres are frozen in time and in the compositional rules they must conform to. So why is it that contemporary Jazz is so different from the Jazz we used to play in the 50's? Precisely because Jazz is not a musical genre *per se,* nor is it a collection of sub-genres.

Taking its roots more than a hundred years ago in Blues, Gospel, and Ragtime to name only a few, Jazz underwent an amazing number of transformations, leading to trends such as New Orleans, Swing, Gypsy Jazz, Bebop, Cool Jazz, Free Jazz, Hard Bop, Modal Jazz, Jazz-Rock, Fusion, Latin Jazz *etc.* In many cases, those trends appeared out of the will to incorporate some foreign musical element into Jazz. For example, New Orleans borrows from French Antilles, Gypsy Jazz from (French) accordion music, Latin Jazz from Cuban and African rhythms, Jazz-Rock from Rock (obviously) *etc.* The list goes on and on. Within this context, a crucial point to understand is that contemporary Jazz is not "the latest sub-genre we came up with until the next one appears"…

> *that is why I prefer to speak of* trends *rather than of styles or genres*

…Jazz, instead, is more a philosophy of music, even more so a musical *process*: grabbing musical bits and pieces everywhere, and making a happy melting pot out of those influences, in a way which is specific to each Jazz musician. In this way, a Jazz musician adapts the music to his own desires instead of adapting his ideas to the constraints of one particular musical style. Exactly as a Lisp programmer adapts the language to his own needs instead of adapting his ideas to the constraints of the language. Note also that this is in full concordance with the notion of positive transgression: a way to move forward by crossing some barriers without denying them, by constantly pushing the limits away, establishing new sets of rules encompassing a larger universe, always in expansion.

The other angle is that of improvisation, a concept at the core of Jazz since the early days of Afro-American slave music…

> *in fact, I would be extremely reluctant to call "Jazz" any kind of entirely written music, be it in a Jazz style like swing or ballad*

…Improvisers usually develop their discourse based on some characteristics of the score, a context such as the rhythm pattern(s) and harmonic progression(s). Not unlike a speaker may develop an argument around a particular topic. Even if the improviser strictly conforms to the indications written in the score, some form of transgression is already present, as the improviser will at least take a lot of liberty with the theme, playing something entirely different, not on the score at all (that is the difference between an improvisation and a *variation* as found in classical music). But improvisation goes in fact much farther than this, as there is essentially no rule when you improvise.

For as much as learning to improvise is a gradual process in which you progressively relax the constraints, a full blown improvisation can contort the original tune in any possible way. In some sense, improvising lets you modify the score in real-time, simply by deciding to change the ambiance, rhythm, harmony, or whatever you may see fit. A particularly transgressive form of improvisation lies in the notion of "playing out" which, as its name suggests, pretty much boils down to being completely off-topic for a while (see appendix C.2).

All of this shows that Jazz is inherently an extensible musical language. As in Lisp, the Jazz transgressions are only possible because the language not only allows them but supports them by providing transgression tools. With some experience for instance, playing out becomes an integral part of the art of improvisation, a rule in itself rather than an exception. The *licks* developed by Miles Davis and other pioneers are now well known, and you can learn actual techniques for chord substitution and dissonant improvisation in Jazz schools. Just as Lisp provides tools for self-transgression by extensibility, Jazz provides techniques for self-transgression by harmonic or rhythmic alteration.

### 5.3.2 Introspection

Let us emphasize again the notion of introspection. Because Jazz is a philosophy of music rather than a specific style, it is, and will remain, the most introspective musical form ever. A true Jazz musician is not interested in mastering such or such musical style to perfection, but in developing his own musical language through composition and improvisation. Just as Lisp can become very personal for every programmer, there is essentially one Jazz language for each Jazz musician. When he was talking about style in programming, Dijkstra made this connection, although to a lesser extent than what we do here {20}. In Jazz, a teacher of improvisation does not teach what to improvise, but rather helps his students find the means to express what they already have, deep inside of them. In Jazz, we not only seek our own style, but our own complete language. It is in that sense that John Foderaro could have said that Jazz is the "programmable musical language" par excellence {18}, or Alan Kay that Jazz isn't a musical language: "it's building material" {19}.

The fact that there is essentially one Jazz language per Jazz musician probably has the same kind of drawbacks as in the case of Lisp, notably in terms of popularity (the public for Jazz is small compared to that for Pop or Rock). Knowing Lisp may not be enough to understand every Lisp program out there, simply because a Lisp program may be stuffed with *ad hoc* embedded DSL's with non-standard syntax and semantics. In a similar vein, knowing Jazz well is usually not sufficient to fully understand the discourse of every improviser, because every such improviser plays with their own licks, idiomatic expressions *etc.*, in other words, with their own, *ad hoc,* embedded musical DSL's. Thus, in order to fully appreciate the music, one has to get accustomed to the musician — a process requiring time, energy, and commitment (even more so when you're not a musician yourself; it is sometimes said that Jazz is a music for musicians only). Not everyone is willing to do so, perhaps only the more demanding {21}, especially in the leisure area.

## 5.4 Aikido Transgressions

Granted, whatever the martial art, some rules simply cannot be transgressed. Breaking the etiquette (the Reigisaho) is very bad and could lead to exclusion in the early days of the discipline, because these are the rules that guarantee the integrity of "the house" (the Dojo), its peaceful life, and the welfare of its practitioners. Apart from that, any martial art claiming to remain a traditional Budo will make the notion of transgression almost a requirement.

The non-competitive aspect of Aikido ensures that at least in theory, anything can happen just like in a real life and death situation. Any rule can be broken, or essentially, there is no rule. Just like in Lisp or Jazz, the transgressive aspects of Aikido may be viewed from the angles of extensibility and introspection (concrete examples may be found in appendix C.3).

### 5.4.1 Extensibility
The fact that Aikido, being a traditional Budo (art) rather than a Bu Jutsu (technique) or a combat sport, does not focus on the techniques themselves is probably the most difficult aspect of the discipline to understand from the outside. This misunderstanding triggers too many and too frequent questions…

> *Is Aikido efficient in a street fight? Does it beat Karate?*

…All these questions are, in fact, irrelevant.

Even though in theory, there are no rules or if there are, they can be broken, we obviously observe rules when we practice. We use a set of codified techniques to improve our mastery of the art, just like a musician practices scales or fingerings in order to improve his mastery of the instrument. However, we always keep in mind that the techniques are only means to work on something more profound.

This has an important consequence. Because a particular technique is a means rather than an end in itself, it could in theory be replaced with something else, or evolve over the years. This is why Aikido is inherently extensible, something which has been stressed over and over again across its history. The founder himself explained this very early and on several occasions [92, 93]. His son did as well {22}. Indeed, the contemporary Aikido landscape is quite diverse. Today, there are numerous schools and variants, not unlike the different Jazz trends or Lisp dialects. Because Lisp is a programming language philosophy rather than a particular language, because Jazz is a music philosophy rather than a specific musical form, because Aikido is a Budo rather than just a Bu Jutsu or a combat sport, these disciplines can all be expressed in many different and evolving ways, without losing their integrity.

### 5.4.2 Introspection

Unsurprisingly, introspection is a central concept in Aikido as well. As a traditional Budo, the whole purpose of Aikido, through martial techniques, is in fact personal development. Aikido puts its practitioners in martial situations, and asks for peaceful resolution of the conflicts through the unthwarted circulation of the energies. This has two important implications, one physical and one mental. On the physical plane, letting the energies circulate freely requires a deep understanding and mastery of your own body, its balance, its position in space, and its interaction with your partner. On the mental plane, peacefully resolving the martial conflict requires a deep understanding and mastery of your emotional state, in order to free yourself from anger, fear, or any other emotion likely to deprive you from your flexibility and suppleness.

From this, it follows quite naturally that Aikido, just like Lisp or Jazz, is in fact very personal, as we all have different bodies and mental postures. Aikido is thus a body language: an opportunity to work on ourselves through a physical medium. Along with Dijkstra {20} and Jazz teachers, an Aikido teacher does not teach Aikido, but rather ways for a student to find his own Aikido. This has been strongly emphasized across the whole history of the discipline, starting in 1935, when Morihei Ueshiba wrote down some Aikido guidelines for the first time (breaking the tradition of oral transmission) {23}, and on numerous other occasions {24,25}. One particular interpretation of the term Sensei*, used to address a teacher, also confirms this: "someone more advanced than you on the way". In other words, a Sensei merely indicates a direction, but doesn't tell you how to follow it.

### 5.5 Transgression, Change, Daredevils

People feel differently about transgression. Some live for it, some hate it. Whether one likes or dislikes transgression is probably connected to either the taste for risk, or the fear of the insecurity that change procures.

Take the case of Common Lisp's `loop` macro for example. As noted by Peter Seibel [78, chapter 7], `loop` is controversial in the Lisp community. Some people love it, some people hate it. I think it is to be expected because `loop` is a very transgressive construct.

People usually hate it because it is considered very unlispy (infix syntax, reserved keywords *etc.*). On the contrary, the very fact that such a thing as `loop` can be written is extremely lispy in my opinion. It is very transgressive of regular Lisp, but it is totally in the philosophy of the language to let us write such things, and this is precisely what I love about it.

In the same way that some people hate the `loop` macro, some people also hate the "next Jazz trend". In their own respective times, new trends such as Bebop, Free Jazz, and Fusion (notably the electric period of Miles Davis) were heavily criticized by Jazz "purists" [33]. The introduction of frictions and dissonance in improvisation suffered the same fate. Every single time this happened, I believe that the offended were systematically seeing Jazz as a specific style of music rather than as a philosophy of music, or a musical process. Their bad.

*There is fun in transgression. But looking at a single system, there can be only so many acts of conformance or transgression before you end up knowing it from the inside out, and lose interest. Perhaps, then, the time has come to play with other systems…*

## 6 Aesthetics of Unification

Until now, we have been considering systems of rules one at a time. We now want to consider multiple systems at the same time, and explore the latitude that multiplicity leaves us.

At first glance, it would seem that being governed by several systems of rules at the same time can only mean more constraints. There is however one way to regain freedom: by succeeding in combining those systems back into a single one. There is enlightenment in drawing bridges between potentially unrelated fields, starting to figure out what is the common essence of things (it is, after all, the whole point of this essay). This is what we call the *aesthetics of unification*.

### 6.1 Epistemophilia

*Epistemophilia* is the pleasure of learning new things. Knowledge acquisition can indeed be an immense source of pleasure. I see myself as an eternal student, and the pleasure increases when I take on new areas. Another level of pleasure is reached when learning about new areas allows me to make new connections with fields I already know. The pleasure of unification is thus deeply connected to epistemophilia, which may be explained with a little bit of epistemology.

As noted by scientists and philosophers alike, understanding the universe we live in has always been an urge for humanity {26,27,28}. Science, as well as belief systems such as religion or myths, are devoted to that purpose. The critical difference between Science and belief systems, however, is that scientific theories need to be validated by experimentation, and experimentation cannot be done on the universe as a whole.

Thus, in order to progress, Science first needs to split the universe into small, "experimentable" bits, which leads to parcelling {29}. It is only afterwards that another form of progress may be seen: reuniting what was broken into pieces before {30,31}.

Unification may involve domains more or less distant from each other, and I can see three major variations of the concept. One, that we may call *homogeneous unification* would consist in bridging gaps within a single discipline. In physics for example, the so-called GUT (Grand Unified Theories) and ultimately the TOE (Theory of Everything) are attempts at unifying the different forces.

Slightly more transversal, another form that we may call *mixed unification* would consist in bridging different scientific domains together, either by analogy (a weak form of unification), imitation, or by exhibiting stronger connections, notably behavioral. Computer science has seen this happening with biology on several occasions, with Alan Kay's early view of object-orientation [52], Artificial Intelligence in general (neural networks in particular), the study of complex systems [82, 107], and perhaps even the early work of Lehman on software evolution [62]. Recently, I tried to demonstrate how very biological the tinkering aspect of software evolution is in some systems [97, 98]).

In a much broader and transversal scope, *heterogeneous unification* would consist in drawing bridges between completely unrelated disciplines, not even all scientific. Perhaps the most striking example of this in computer science is the concept of *design pattern* [18, 19, 20, 40, 56, 77] originating in the field of Architecture [2]. Again, note that attempting to draw bridges between computer science, music, and martial arts as done in this very essay is probably as transversal and heterogeneous as it may get.

It seems that the pleasure of unification increases with the distance separating the unified domains. This is probably related to the difficulty of the process: the greater the challenge, the greater the reward.

## 6.2 Apophenia

Be it extreme cases of transversality or incongruous unification attempts (e.g. considering a LaTeX document as a unicellular organism [98]), one is bound to ask whether the perceived connections are delusional, or make any actual sense. Do we discover them or do we invent them? Once discovered or invented, do they become useful or do they remain a sterile intellectual construction? In this context, an interesting concept is *apophenia*. Originally coined by psychiatrist Klaus Conrad [21], apophenia can roughly be defined as a human tendency to see or seek order in randomness, or by extension, common patterns in unrelated contexts. Thus, we can see very clearly how apophenia is related to unification. In her analysis of apophenia [50], Hubscher notes the importance of this trait in evolution: by seeing danger patterns, even where there is potentially none, a population is led to adopt a "better safe than sorry" attitude, hence increasing its chances of survival at the risk of being overly cautious. She also stresses the link between apophenia and creativity, exhibited by several neurobiologists.

For Heilman [46], creativity is seen as "the ability to understand and express novel orderly relationships" (unification, in other words). Brugger [16] makes two other points. First, the ability to make connections and especially remote rather than close ones is at the heart of creative thinking. Next, Brugger establishes a gradation in apophenia, with detection of actual patterns at one end (creativity) and interpretation of noise as patterns at the other end (hyper-creativity, delusion).

For people like me, who tend to think that randomness doesn't exist (being just a buffer for things we don't know how to model yet), the thought of hyper-creativity and delusion is a bit worrying. As concluded by Hubscher, however, even the highest degree of apophenia is not necessarily a negative trait {32}.

Finally — and back to the notion of enlightenment that unification provides (a sense of new understanding) — the question of delusion is probably of little importance. It doesn't matter if the result of unification is an epiphany or an *apophany*. A new scientific theory is neither true nor false. It is just a model that works or doesn't work in a particular context, no matter how the theory comes to birth. In Art, the result of a creative process doesn't even have to "work" in the scientific sense. As a consequence, the only thing that matters is that, fueled by various degrees of apophenia, unification is one facet of creativity. The enlightenment that unification provides, whether delusional or not, is always a source of pleasure.

### 6.3 Unification in Lisp, Jazz and Aikido

Each in their own respective domain, Lisp, Jazz, and Aikido share a common concern for unification.

#### 6.3.1 Lisp

The first unifying aspect of Lisp lies in the fact that it is a multi-paradigm language. Today, Lisp offers an extremely wide range of such paradigms, ranging from low level ones like imperative and procedural to more abstract such as object-oriented and functional. Of course, many younger languages today would qualify as multi-paradigm as well. However, Lisp goes further (and has done so for a longer time) than most other, more "modern" languages, probably because of its original design and philosophy. Several examples of this come to mind.

First of all, many paradigms in Lisp are available, not necessarily because people felt the need for them, but simply for the sake of the exercise: because it was possible, fun, and easy to do (or easier than in other languages). For instance, it is notoriously simple to add logic programming (Prolog) to Lisp [71, chapter 11] [42, chapter 24]. Similarly, ContextL [22] was the first full implementation of the context-oriented programming paradigm [47].

Another aspect of the multi-paradigm philosophy of Lisp is to remain paradigm-neutral. This is especially true of Common Lisp, which adopts a pragmatic attitude with respect to programming style, sometimes sacrificing a bit of cleanness or theoretical purity for the sake of being more practical. While being very high level and abstract, Lisp also lets you perform bit-wise operations and use `goto` statements, in spite of all the controversy around it [27, 58]. Contrary to Haskell, Lisp doesn't put side-effects into quarantine. Contrary to Smalltalk, Lisp didn't become full-object once CLOS was added to it, and CLOS won't force you to be message-passing exclusively. The unhygienic macro system lets you capture free variables if you wish to do so, and even *a* *priori* antagonistic paradigms are supported simultaneously, such as lexical / dynamic scoping, or static / dynamic typing. In other words, Lisp is not judgmental about paradigms. In the same way as there is no wrong note in Jazz, there is no wrong paradigm in Lisp.

Yet another aspect of unification lies in the software life cycle. Contrary to most mainstream languages, Lisp unifies the development, compilation, execution, and debugging phases. Even if one wishes to generate a standalone executable, it is possible (even the default) to embed a REPL (Read-Eval-Print Loop), compiler, and debugger in it.

### 6.3.2 Jazz

The unifying aspects of Jazz are extremely similar to that of Lisp. First of all, Jazz, as a philosophy of music rather than as a particular style, is a multi-paradigm musical language par excellence. This was twice summarized brilliantly by famous French pianist Michel Petrucciani {33,34}. Just like Lisp, Jazz encompasses every possible musical paradigm (style), not by necessity, but simply because it is possible to do so. A Jazz musician is an explorer before anything else, interested in all musical forms, ranging from traditional swing to binary or Latin grooves. Jazz musicians seldom restrain themselves to a specific genre, and usually hate to be categorized.

Jazz is also like the Common Lisp of music, pragmatic rather than pure or clean, neutral about the style. Jazz will never force a single-paradigm approach. A Jazz musician will never have to remain purely swing or full-Bossa-Nova. While the Keith Jarrett trio has the beauty and cleanness of a purely functional program, the heavy-metal sound of Tribal Tech feels more like an unhygienic macro call or a `goto` statement. Yet, all this is possible and Jazz is not judgmental about it. Also, just as Lisp supports antagonistic paradigms which you are free to use at the same time, Jazz lets you do that as well. Perhaps the most typical illustration of this is the Jazz-Rock / Fusion trend which started in the early 70's. For example, listen to Tribal Tech's Susie's Dingsbums [85, track 6] or Soul Bop's live interpretation of the Brecker Brothers' Above and Below [4, disk 1, track 3], and enjoy the `longjmp`'s from traditional ternary swing to either Latin or Rock-like binary grooves within the same song…

Finally, another similarity with Lisp lies in the music life cycle. Lisp unifies the development, compilation, and execution phases of a program, and Jazz does exactly the same with music, notably through improvisation and the extreme freedom of interpretation it allows. I have always considered that a composition is just an improvisation that happens to be fixed in a score, and that conversely, an improvisation is a real-time composition, forgotten as soon as it has been executed. In a functional sense, an improvisation is like an anonymous expression, and a composition a binding to it. Through improvisation, Jazz literally turns your band into a REPL, and gives you the ability to modify the song (the original program) in real time, essentially mixing the musical development, compilation, and execution phases all together. But note that debugging is included as well. Because Jazz is so forgiving (fault-tolerant) about your mistakes, and because improvisation is a very risky business, an improviser will often make errors and correct them on the fly. It is part of the game. Listen for example to how Michael Brecker throws a `high-harmonics-error` on his saxophone at the very beginning of an improvisation, catches it, debugs it, and restarts the phrase, all of this in real-time [15, track 1]. Surely, it takes at least something as powerful as the Common Lisp condition system to be able to do that [78, chapter 19]. It also reminds me a lot of how Graham used to debug the Yahoo! Store in real time [41, chapter 12]…

### 6.3.3 Aikido

Linking the unifying aspects of Aikido to those of Lisp and Jazz is probably a bit more delicate, as the similarities are more theoretical than practical. Nevertheless, unification is also at the heart of the discipline, and one might be equally sensitive to it, even though its forms may be slightly different or less apparent. Evidence of the unifying aspect of Aikido comes in at least three important forms.

First of all, Aikido is a multi-paradigm martial art to some extent. It is well known that before Aikido was created, its founder Morihei Ueshiba had been practicing many disciplines, and was notably a master in Ju Jutsu*, Ken Jutsu* and Jutte Jutsu*. The technical part of Aikido is hence the result of mixing together and adapting forms coming from a lot of different horizons. Today, this multi-paradigmatic aspect is more theoretical than practical because regular practice only involves bare-hand techniques and only three different weapons. Remember however that at least in theory, everything is possible in Aikido because the techniques are only a means, not an end. That is why there is nothing in Aikido which would prevent the use of other weapons, kicks, or anything else. For the same reason, Aikido will never enforce a single-paradigm approach either. Contrary to Judo, you may use weapons, and contrary to Iaido* or Kendo* (two weapon-centered Japanese disciplines), you are not forced to use a single kind of weapon. Aikido is also paradigm-neutral in the sense that practicing with a beginner doing a lot of mistakes (for example using unhygienic grasps or low-level jumps) is valued as much as practicing with a purely functioning expert.

A second form of unification in Aikido is related to the very creation of the discipline: the famous "1925 revelation" of its founder. At that time, the martial skills of Morihei Ueshiba were growing in fame and many of martial artists wanted to fight him, in order to see for themselves. After defeating yet another opponent, a Kendo master, almost without fighting (only by eluding), he had the sudden revelation that his art should be an art of peace, preserving life instead of destroying it, and felt a very strong connection with the universe as a whole. This is how the discipline went from *Aikijutsu* (technical) to *Aikibudo* (a philosophy), and eventually to *Aikido* (The Way). On a spiritual plane, the founder's interpretation of the notion of Budo hence not only encompassed the traditional meaning of the term (notably including personal development) but also extended to such notions as love, protection of all things, and respect for all lives, which are much more transversal (universal) concerns. Since then, the founder never ceased to emphasize the importance of unification and harmonization {35}, with words amazingly close to those of Boole {36}.

Given the highly mystical personality of the founder and his writings, whether his 1925 revelation was an epiphany or an apophany is left to the reader's opinion…

> *perhaps it is safer to simply use the Japanese term for this: Satori**

…Anyway, the result is that many martial art masters at the time considered Morihei Ueshiba's work as the ideal Budo, encompassing all of the previous ones. This is exactly what Jigoro Kano, the founder of Judo said, when he discovered it. He later sent his best students to Ueshiba and some of them like Kenshiro Abbe (Japan Judo champion) stayed for 10 years with him [75].

Finally, although not specific to Aikido but rather to any non-competitive physical activity, the third important form of unification we want to exhibit is related to the discipline's life cycle. Because there is no competition, practice is not split into separate phases such as training (development), competing (execution), and debriefing (debugging). Instead, an Aikido partner is just like a REPL or your Jazz band, letting you experiment, improve, make mistakes, fix them, all of this at once, permanently, and in real time. Just like Lisp and Jazz, Aikido is extremely interactive and dynamic in nature. Even preparing or passing grades is normally not considered as separate phases in Aikido. When you pass a grade, you simply do your best and try to practice as usual.

## 6.4  Minimalism

One very important aspect of unification, shared by Lisp, Jazz, and Aikido is a concern for minimalism…

> *minimalism can be seen as a process leading to simplicity*

…We have already tackled some aspects of simplicity (in section 4.2.3). Let us only revisit the matter from the angle of unification. There are essentially two ways of unifying systems of rules. One is creating a more complicated new system, encompassing the others, for example by merely performing the union of the previous rules. The opposite approach, however, is to find more general, more expressive rules, in which the former ones become special cases of the new. In doing so, the number of necessary rules is reduced instead of increased. From an aesthetic perspective, this approach is much more pleasing because it leads to a "win-win" situation: not only is the system is simplified, it becomes more expressive.

### 6.4.1  Simplification

One key to unification in Lisp, Jazz, and Aikido lies in the multi-paradigmatic aspect of the disciplines. But consider again how this happens concretely. In mainstream programming languages, every paradigm usually comes with its own specific syntax. In C++ or Java, Object-oriented programming requires new reserved keywords (`class`, `virtual`, `private` *etc.*). In PHP, variables are prefixed with a $ sign and instance ones with an @ character in Ruby. In Go, one may omit the type declaration of an initialized variable, but the syntax is different (`:=` instead of `var`). Most languages adopt the usual $f(x, y)$ mathematical notation for function calls, yet arithmetic operators are expressed with infix notation. In C++ again, meta-programming is done with the template system, which is completely different from raw C++. The list goes on and on…

In Lisp, on the other hand, new paradigms are made available through existing core constructs, instead of by adding new ones. In particular, almost every paradigm involves only function / macro calls. Structures are created by calling `defstruct`, classes are created by calling `defclass` *etc.* Even functions and variables are created this way. Method calls obey the same syntax as function calls. When there is syntax in Lisp, it is only sugar, and it is customizable. As a result, new paradigms are usually provided as libraries, and the core language doesn't grow. In Racket, a recent Lisp dialect descending from Scheme, this idea is pushed even further, up to the claim that not only paradigms, but complete *languages* can be implemented as libraries [88].

Unsurprisingly, the evolution and growth of Jazz is extremely similar to that of Lisp. As mentioned before, there is not much in the core Jazz language: most of the time, a theme, harmonic progression, and general style indication will suffice. However, just as the minimalist core of Lisp doesn't make it less expressive (on the contrary), the minimalist core of Jazz gives the musician more freedom, not less. When Muddy Waters plays some good ol' Mississippi Blues with almost only the 5 notes of his pentatonic scale, on a not even perfectly tuned guitar, he is no less expressive than any other musician. The range of conveyable emotions is the same, it all depends on what the musician does with his notes. In a similar vein, modal Jazz (in which a song's harmonic progression is reduced, sometimes to a single chord) does not limit the improviser's expressive power, but on the contrary broadens it by removing harmonic constraints.

Over the years, the Jazz landscape did grow a lot, but the core language did not. Every specific Jazz trend such as swing, bop, or Latin, can be seen as a musical *library* that the composer / player is free to incorporate into his musical program, rather than as a new core feature. Just because one day we started to use Latin grooves doesn't mean that all of a sudden, new rules were established to specify how to use them gracefully with the rest of Jazz. A composer remains free to use and mix together whatever styles he wants, just as the Lisp programmer remains free to mix every possible paradigm or (embedded) DSL within the same application.

Looking back at the last 80 years of Aikido, it is obvious that its evolution follows the same pattern. Even though Aikido is difficult, it is not complex, and its complexity has not increased over the years. Aikido doesn't have hundreds more techniques or weapons every year. As mentioned before, the core of Aikido is as small as that of Lisp or Jazz: the only fundamental rule is to let the energy flow naturally and in suppleness. Yet, the founder himself taught that the number of techniques is infinite and they evolve constantly {22}. How is this possible? Simply because Aikido techniques do not belong to the core, but are more like libraries of moves which you are free (not) to use. As experience grows, an Aikido practitioner realizes more and more deeply that all the techniques are essentially the same: ways to achieve the same goal, differing in surface only. When faced with an attack such as Yokomen Uchi*, it doesn't really matter whether you use Shiho Nage*, Kote Gaeshi* or any other technique to immobilize your partner. These are just different paradigms in a Ki-complete language, just like computing a factorial iteratively or recursively will get you the same result (there is no Ki overflow in Aikido).

Incidentally, this particular kind of growth makes the evolution of Lisp, Jazz, and Aikido very close to that of natural languages. The complexity of natural languages does not increase over time: while their grammar evolves very slowly, most of the novelty comes from the addition of neologisms and new idiomatic expressions. Just like natural languages, Lisp, Jazz, and Aikido grow by "vocabulibrary" [1, 100]. This particular trait may explain, at least in part, their longevity.

### 6.4.2 Meta-Levels and Homoiconicity

Unifying systems by simplification and reduction of rules has one very important consequence, equally shared by Lisp, Jazz, and Aikido. The core of a minimalist system necessarily only expresses the essence of the corresponding domain, along with a meta-level making the core expandable in every possible direction. The core of Lisp is the lambda calculus, the core of Jazz is improvisation, and that of Aikido is the circulation of the Ki. By seeing every possible musical style, every potential martial technique as a means to study their respective essence, Jazz and Aikido become meta-disciplines, just as Lisp is meta-programmable. Adopt the Jazz attitude and the song you play automatically becomes Jazz. Take the posture of an Aikidoka and your movement automatically becomes Aikido. In other words, Jazz and Aikido are defined by the simple fact of deciding that they are. Just like Lisp, Jazz and Aikido are in fact *meta-circular*.

In Lisp, meta-programming is achieved through reflexivity and extensibility. What makes this not only possible, but *simple*, however, is in fact the homoiconicity of the language [53, 68], a fundamental trait of Lisp, later shared by only a few other languages such as Prolog [76]...

> *homoiconicity simplifies the simplification process*

...In a homoiconic language, the program's denotation corresponds exactly to its internal representation (hence the expression "code as data"), in a user-accessible data structure. That is why meta-programming is so simple and homogeneous (one ends up manipulating data which happens to be code). Homoiconicity is probably the ultimate form of unification one can encounter in a programming language, as it essentially removes the deepest and most fundamental frontier: that which lies between the language itself and its usage. The language, the program, and the programmer essentially become one.

A similar phenomenon can be observed in music. Many musicians experience a depressing phase during which we don't seem to improve anymore. Then all of a sudden, we pass a mysterious threshold. Only afterward we realize that the instrument has become an extension of our own body, and that we have "interned" (a lispy way to say "internalized") the music. The musician, the instrument, and the music become one. Perhaps this explains why the voice occupies such a special place in the musical landscape: the instrument is not even an extension of the musician. It *is* the musician. Perhaps this also explains why Bobby McFerrin — unfortunately known for his least interesting piece: Don't Worry, Be Happy — has been in my "desert island picks" forever. The way he uses his voice like many instruments at the same time [66, 67] makes my homoiconic fiber resonate as no-one else does.

Aikidoka endure the same kind of roaming in their progression. At some point, we must feel the weapon as an extension of the body and "intern" its moves. In Aikido the purpose of weapons is only to improve bare hands techniques, and genuine homoiconicity is reached in the Shuto*: the techniques in which the hand itself serves as a saber. The discipline, the weapon, and the practitioner become one.

This drastic reduction in distance is also deeply related to the mastery of the Art. A complete mastery of any language is reached when you basically forget all about it, and express yourself naturally with it. When we speak, we don't think about the grammar of our mother tongue anymore. Instead, we express ourselves directly. When we improvise, we don't think about the scales, the keys, or the instrument anymore. Instead, we express our musical self spontaneously. Now consider how similar the whole point of Aikido is [90]: "Be it in the artistic or cultural domains, or in the Budo, it is necessary to practice in order to master the form, but once mastered, one must free oneself from it." In Aikido, the ultimate training exercise for this is the Randori*. When attacked by four partners at the same time, you don't have time to think. Just like an improviser on stage, you only need to act {37,38}.

### 6.4.3 Minimalism, Dynamicity, Daredevils

As easy as it may be, what is absent from the core of a minimalist system still needs to be reconstructed at some point. The freedom granted by the meta-level, however, is to be able to reconstruct *only* what is needed, *when* it is needed. Such systems are hence extremely dynamic in nature, and as for transgression and change, people feel differently about minimalism and dynamicity.

Programming in a highly dynamic language such as Lisp is indeed a risky business (fewer safeguards, if at all, than in a statically typed language for instance). Jazz improvisation is an extremely risky business as well. So is putting yourself in martial conflicts situations.

The most frightening instance of dynamicity and minimalism in Jazz is without a doubt the Free Jazz trend from the 50's, pioneered by musicians such as Ornette Coleman and Cecil Taylor, and in which the only rule is that there is none. Free Jazz is also the most misunderstood Jazz trend, as many people (musicians included) mistook "do what you want" for "do nonsense". On the contrary, the whole purpose of Free Jazz was not to produce random noise, but to construct something coherent from scratch (from either a very short melody, or nothing at all). In doing so, Free Jazz musicians would often actually construct their own musical rules at the same time they were using them.

Note how creating live music from void is very much like writing a program from an empty REPL, not even having a skeleton of a `main` function to start with. Along the same lines, creating your own set of musical rules at the same time as using them is not unlike a Lisp programmer shaping the language itself (with intercession, extensions, or embedded DSL's) at the same time the program develops.

Free Jazz in Aikido is called Jiu Waza*. It is customary to end every practice session with this form of work in which the practitioners are free to use any kind of technique they want. You start from a clean slate and you let something grow out of it. Just like an empty REPL or a blank score, Jiu Waza can be a very a scary exercise, especially for beginners. Not having a specific move to execute is already frightening, although much less than not even knowing how you are going to be attacked…

### 6.4.4 Uniformity

People interested in Lisp, Jazz, and Aikido enjoy going meta. They enjoy diving below the surface, very deep, to the core, to the essence of things. And at that level, everything looks basically the same.

Look at appendix F and notice how smooth, fluid, and homogeneous the Lisp code is, notably compared to the C++ code below it (even at the glyph level, parentheses are smoother than curly braces…). I believe that even the people complaining about the amount of parentheses in Lisp are in fact not so much complaining about that, but instead about the fact that they see parentheses *only*. By comparison, C-like languages are ragged, and it is supposed to help discriminate the contents of the code. While Lisp code can look completely uniform, it doesn't mean that there is less variety in it. It's just that the variety is less apparent, and that consequently, it takes a deeper knowledge of the language and / or the libraries to discern and segregate the various constructs.

Jazz and improvisation often sound equally uniform, particularly to novice ears. There is an incredible wealth of harmonic diversity and cleverness in the Jazz standards for instance. However, because most of them follow the usual rhythmic setup (ternary groove, walking bass *etc.*), they all have a tendency to sound extremely similar if you don't pay attention. Discerning the subtleties in harmonic progressions takes some learning and getting used to. The same goes for improvisation. A fast bebop improvisation may give the impression that the musician plays random notes, only very rapidly. This, of course, is completely wrong. There again, a minimum of cultural background or even better, knowledge of the particular musician is required to fully understand the connection between his lines and the underlying composition, his own vocabulary, favorite licks, and idiomatic expressions. In spite of its richness and complexity, the apparent uniformity of Jazz is what makes it very well suited to serve as background (but undisturbing) noise in elevators, which is both a good thing and a shame…

Finally, Aikido also has this property. First of all, and mainly because it is all about the unthwarted circulation of energy, Aikido is extremely homogeneous, fluid, and smooth. There are no bumps or lumps in Aikido. Only roundness, like parentheses, not curly braces. A lot of people seeing it for the first time think that it looks like a dance. But Aikido also looks very uniform because most of what happens can only be felt from the inside, and is not visually apparent. Learning by watching (Mitori Geiko*) has its limitations. Only with at least minimal knowledge of the moves will the spectator be able to appreciate what's really going on. This is especially true when looking at the masters. When you reach a certain control of the flow of energy, the techniques practically disappear, the moves are reduced to their simplest, purest form, and you see nothing more than a mass displacement rushing through its partner.

Just like an empty REPL, a blank score, or an unknown attack, uniformity may either frighten or fascinate. For me, it is the latter. The aesthetics of uniformity is directly linked to the pleasure there is in being able to figure out what is hidden

Didier Verna

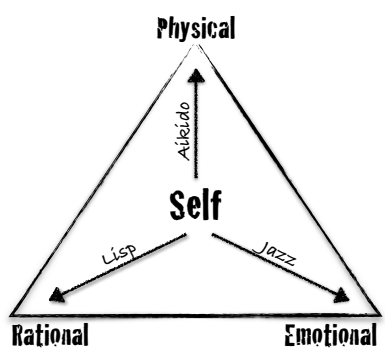 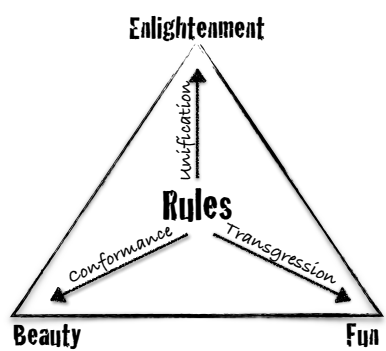

**(a)** The Triangle of Languages  **(b)** The Triangle of Pleasures

**■ Figure 1** The Triangles of Life

behind it, not unlike Neo finally being able to "see the Matrix". By discerning shapes in uniformity, one essentially crosses the "Final Frontier" (perhaps the ultimate form of transgression, and the most difficult one), accessing the essence of things. This is a form of transcendence and the pleasure it gives is that of sublimation.

## 7  Conclusion

It is easy to forget the importance and scope of the word "language" in the expression "programming language". Whether programming or otherwise, we have a very intimate relationship with the languages we learn and use, and this relationship goes both ways. As means of expression, the languages we choose should be very close to who we are and what we want to say. As means of reflection, the languages we use also shape our minds, hence changing who we are and what we want to say.

The importance of creativity and the artistic (hence emotional) nature of the act of programming has long been acknowledged. We seek pleasure when we program, and we know that an important driving force to achieve it is a sense of aesthetics. However, perhaps this highly psychological aspect of programming has not been explored to its full extent. By placing the notion of aesthetics (as a driving force) at the center of the debate, we may be able to answer a question which has not been studied much until now: absent any industrial constraint or necessity, what motivates the choice of a programming language? Why and how does one become our favorite? In other words, can we explain our "desert island pick"?

As human beings, we are intrinsically multi-modal, so we need different languages to express ourselves and know ourselves better through various communication channels. Three dimensions strike me as being of the essence here: we are physical beings, we have a rational mind, and also an emotional one. As an individual, my profound taste for introspection and self-development led me to cultivate with equal importance those "three angles of self" and, over the years, I discovered a "triangle of languages" (figure 1a) each of which made me feel at home. Lisp is a language for the rational mind, Jazz is a language for the emotional one, Aikido is a body language.

Why those three in particular? Again, it is possible to answer this question by focusing on their creative nature. A fundamental motivation in both programming and Art (whether martial or otherwise) is seeking pleasure by exercising a set of aesthetic dimensions, and as I was discovering my personal triangle of languages, I also eventually discovered a "triangle of pleasures" (figure 1b) each of which made me feel equally at home. At the center of it is the notion of systems and rules. From this center, one may adopt three different postures. Conformance is the act of obeying the rules, and the pleasure it gives is that of achieving beauty. Transgression is the act of breaking the rules, and the pleasure it gives is that of having fun. Unification is the act of merging the rules, and the pleasure it gives is that of enlightenment.

Why does the triangle of languages map so well on the triangle of pleasures? Simply because Lisp, Jazz, and Aikido share the exact same view of conformance, transgression, and unification. One can in fact arguably claim that these languages are just three expressions of a single aesthetic essence, only on three different dimensions. In terms of conformance, Lisp, Jazz, and Aikido strive for providing only a few simple rules with a strong sense of support and collaboration. In terms of transgression, they emphasize the idea that a language is not a frozen point of arrival, but on the contrary, a flexible starting point from which to develop a deeply personal variant. In terms of unification, they aim at being as universal as possible, not by accumulating complexity but on the contrary by only expressing the essence of things in a smooth, fluid, and uniform way, along with a meta-level to ensure their potential expansion.

By looking at the triangle of pleasures and how Lisp, Jazz, and Aikido envision these aesthetic dimensions, one cannot help but notice the similarities with the so-called "MIT / Stanford style of design", *a.k.a.* the "MIT approach", or even just "the right thing" [38]. This is not surprising since Lisp actually comes from there. Gabriel also mentions that the opposite philosophy, "worse is better", has strongest survival characteristics. With 60 years of Lisp life, 80 years of Aikido, and more than a century of Jazz, perhaps we may soften this statement a bit, and instead suggest that "right things", being closer to utopias than to realities [14], are simply condemned to remain niches.

While this essay has focused on the notions of systems and rules, much more could be said on the subject. For example, Lisp, Jazz, and Aikido are very modern without being recent, very dynamic and interactive in nature, although we only scratched the surface of these aspect. The exploration of their common essence may thus continue…

**Acknowledgements**   I am particularly grateful to Richard Gabriel for his shepherding, time, support, comments, suggestions, and in general for the exchanges we had during the finalization of this essay.

## Quoted References

{1}  *Science is what we can explain to the computer. Art is everything else.* —Donald E. Knuth [59, p. 168]

{2}  *When I speak about computer programming as an art, I am thinking primarily of it as an art form, in an aesthetic sense. The chief goal of my work as educator and author is to help people learn how to write beautiful programs.* —Donald E. Knuth [57, p. 670]

{3}  *It is my purpose to transmit the importance of good taste and style in programming, the specific elements of style [serving] only to illustrate what benefits can be derived from style in general.* —Edsger Dijkstra [28, p. 6]

{4}  *An understanding, a feeling for the aesthetic of programming, is needed, and not only as a driving force for the programmer [...]* —Andrei Ershov [34, p. 502]

{5}  *The ultimate usefulness is to create a source of pleasure.* —Jeremy Bentham [7, p. 205]

{6}  *It's not true that necessity is the only mother or father of invention.* —Donald E. Knuth [79, p. 100]

{7}  *Whatever I do, I try to do it in a way that has some elegance; I try to create something that I think is beautiful.* —Donald E. Knuth [59, p. 91]

{8}  *[...] transversality is non-categorical and non-judgemental. It defies disciplinary categories and resists hierarchies. A transversal line cuts diagonally through previously separated parallel lines [...]* —Helen Palmer & Stanimir Panayotov [73]

{9}  *A mathematician, like a painter or poet, is a maker of patterns. The patterns, like those of the painter or poet, must be beautiful. There is no permanent place in the world for ugly mathematics.* —Godfrey H. Hardy [45]

{10}  *While the majority of computer scientists have devoted their attention to solving complex problems and introducing new technology, few have written on the thought process behind their creations.* —Ronald J. Leach and Caprice A. Ayers [61]

{11}  *[TeX is] A wildly inconsistent mishmash and hotchpotch of ad hoc primitives and algorithmic solutions without noticeable streamlining and general concepts. A thing like a pervasive design or elegance is conspicuously absent. You can beat it around to make it fit most purposes, and even some typesetting purposes, but that is not perfection.* —David Kastrup [source unknown]

{12}  *An error in the performance of classical music occurs when the performer plays a note that is not written on the page. In musical genres that are not notated so closely [...], there are no wrong notes – only notes that are more or less appropriate to the performance.* —Alan Blackwell and Nick Collins [10]

{13}  *Working with Miles Davis, I played something that was technically wrong. I thought I had just destroyed everything! Miles played some notes, and he made my chords right.*

*I judged what I had played. Miles didn't. That's what every Jazz musician should do.* —Herbie Hancock [44]

{14} *A true warrior is invincible because he fights with no one.* —Morihei Ueshiba [92]

{15} *A good scientist [...] does not merely ignore conventional wisdom, but makes a special effort to break it. Scientists go looking for trouble.* —Paul Graham [41, p. 43]

{16} *[...] the two senses of "hack" are also connected. Ugly and imaginative solutions have something in common: they both break the rules.* —Paul Graham [41, p. 50]

{17} *The important thing is that you really like the style you are using; it should be the best way you prefer to express yourself.* —Donald E. Knuth [57, p. 670]

{18} *Lisp is a programmable programming language.* —John Foderaro [35]

{19} *Lisp isn't a language, it's building material.* —Alan Kay [source unknown]

{20} *I feel akin to the teacher of composition in a conservatory: he does not teach his pupils how to compose a particular symphony, he must help his pupils to find their own style and must explain to them what is implied by this. (It has been this analogy that made me talk about the art of programming.).* —Edsger Dijkstra [28, p. 8]

{21} *If poetry and music deserve to be preferred before a game of push-pin, it must be because they are calculated to gratify those individuals who are most difficult to be pleased.* —Jeremy Bentham [7, Book III, chapter 1, p. 207]

{22} *The practitioner will discover that the techniques are infinite. [...] Rather than following a fixed form, [Aikido ] moves are developed from a common basic principle. It is for that reason that new techniques still appear today.* —Kisshomaru Ueshiba [90]

{23} *The teacher only transmits a limited aspect of the art. Its multiple applications will need to be discovered by every practitioner.* —Morihei Ueshiba [90]

{24} *The Art of Peace begins with you. Work on yourselves.* —Morihei Ueshiba [92]

{25} *Everyone must understand [Aikido ] in a way that is personal.* —Morihei Ueshiba [93, p. 17]

{26} *Humans are creatures that crave to find order and meaning in their environment. Not only do we want to find meaning in our surroundings, but we need to do this.* —Daniel C. Dennett [24]

{27} *It is a requirement to the human brain to put order in the universe. [...] One may disagree with explanatory systems offered by myths or magic, but one cannot deny them coherence.* —François Jacob [51]

{28} *The heart of the problem is always to explain the complicated visible by some simple invisible.* —Jean Perrin [74]

{29} *The beginning of modern science can be dated from the time when such general questions as "How was the universe created?" [...] were replaced by such limited* *questions as "How does a stone fall?". Scientific knowledge thus appears to consist of isolated islands.* —François Jacob  [51]

{30}  *In the history of sciences, important advances often come from bridging the gaps. They result from the recognition that two hitherto separate observations can be viewed from a new angle and seen to represent nothing but different facets of one phenomenon.* —François Jacob  [51]

{31}  *As Science progresses, there is a steady decrease in the number of postulates on which it has to rely for its development.* —Antoine Danchin  [23]

{32}  *Our ability to discern forms in randomness and patterns in chaos is not necessarily a negative trait. From survival strategies of primitive times to the more enriching and impalpable pursuit of beauty and art, we have derived a tremendous benefit by attaching creativity to free association. The entire enterprise of science, after all, is the organized and rational search for order in the seeming randomness surrounding us.* —Sandra L. Hubscher  [50]

{33}  *I'm a real thief. Everyone influenced me.* —Michel Petrucciani  [26]

{34}  *Jazz is like an impolite guest who wants to sit at every table.* —Michel Petrucciani  [30]

{35}  *Use unity to hit multiplicity. […] The universe emerges and shines from a unique source, and we evolve through an optimal process of unification and harmonization.* —Morihei Ueshiba  [92]

{36}  *[…] a perfect method should not only be an efficient one, as respects the accomplishment of the objects for which it is designed, but should in all its parts and processes manifest a certain unity and harmony.* —George Boole  [13]

{37}  *Every encounter is unique, and the appropriate response must emerge naturally.* —Morihei Ueshiba  [92]

{38}  *In Aikido, there is no form or model. The natural movements are the movements of Aikido.* —Morihei Ueshiba  [90, p. 117]

## A   Japanese Terms

**Aikido** (合気道) Literally, the "way (do) of harmonizing (ai) energies (ki)", or even the "art of peace". Japanese martial art founded by Morihei Ueshiba before World War II. Aikido is non-competitive, non-violent, and emphasizes on conflict pacification.

**Aikidoka** (合気道家) Normally, a person professionally involved in Aikido, but outside Japan, the term has been broadened to designate any Aikido practitioner.

**Bu Jutsu** (武術) Literally, "martial (bu) techniques (jutsu)".

**Budo** (武道) Literally, the "martial (bu) way (do)". In other words, the arts of war or martial arts, although in a modern sense, that is, with an educational purpose. This term can probably be attributed to Jigoro Kano, founder of Judo.

**Dan** (段) Grades of the black belt. It is worth noting that the first black belt grade, Sho Dan (初段), translates as the "grade of the beginner". Indeed, in traditional Budo, it is considered that below the black belt, all you can do is learn the techniques. It is only from the first black belt grade that you can start working on the heart of the discipline.

**Dojo** (道場) Literally, "where the way (do) is studied (jo)". The place where martial arts are practiced.

**Iaido** (居合道) Literally, the "way of living in harmony". Iaido is in fact often described as the "art of drawing the saber". More precisely, drawing and striking in a single movement. Iaido is essentially practiced through a set of Kata aiming at perfecting the movements themselves and often considered a good additional practice to Aikido. Notice that the two last ideograms are the same as the first and last in Aikido.

**Jiu Waza** (自由技) Literally, "free (ju) work (waza)".

**Ju Jutsu** (柔術) Literally, "suppleness (ju) techniques (jutsu)".

**Judo** (柔道) Literally, the "way (do) of suppleness (ju)". Japanese martial art founded near the end of the 19$^{th}$ century by Jigoro Kano, which later evolved into a competitive combat sport.

**Jutte Jutsu** (十手術) Literally, "bayonet (jutte) techniques (jutsu)".

**Kaeshi Waza** (返し技) Roughly, inversion or counter techniques.

**Karate** (空手) Literally, "the empty hand". Japanese martial art from Okinawa (in Ryukyu, for a long time an independent province of Japan). Actually imported from China during the 17$^{th}$ century. Its ideographic denotation was changed in 1935 to remove all references to China. The "empty hand" emphasizes bare hand techniques.

**Kata** (方, 形 or 型) In martial arts, a particular form or technique. According to the preferred ideogram, the exact meaning may vary from the "way to do something", the "act of reproducing something", or the "original and ideal form".

**Kendo** (剣道) Literally, the "way of the saber". Can be described as the Japanese form of fencing. Today, Kendo is a combat sport where strikes are blown for real, although most of the time with bamboo sabers. Practitioners, wearing an impressive armor, are easily recognized.

**Ken Jutsu**（剣術）Literally, "saber (ken) techniques (jutsu)".

**Kote Gaeshi**（小手返し）A wristlock-based projection or immobilization technique.

**Kyu**（級）Grades below the black belt. Aikido has 6 Kyu, and contrary to most other martial arts, traditional Aikido does not use different colors for the different Kyu, only white belts. Indeed, the grade is not considered important.

**Meiji**（明治）The Meiji Era (1868–1912) is the period during which the feudal system falls, and Japan switches to an occidental-like industrial one. Literally, Meiji means "enlightened (mei) government (ji)".

**Mitori Geiko**（見取り稽古）Literally, "studying (keiko) by observation (mitoru)". The act of practicing without being physically involved.

**Nin Jutsu**（忍術）Literally, "perseverance (nin) techniques (jutsu)". The set of techniques used by the Ninjas.

**Randori**（乱取り）In Aikido, exercise consisting in being faced with several opponents at the same time (the number increasing with the grade). In such a situation, only simple attacks and techniques can really be used. In Judo, a combat without any stakes, outside the rules of competition, hence in which any technique may be used.

**Reigisaho**（礼儀作法）Roughly, the etiquette.

**Reishiki**（礼式）Roughly, a code of conduct.

**Satori**（悟り）Spiritual awakening, comprehension, understanding.

**Seika Tanden**（臍下丹田）The center of gravity and of all energies, roughly located halfway between the navel and the pubic bone.

**Sensei**（先生）Literally, one who was born before. It has the meaning of teacher, or master in Budo.

**Shiho Nage**（四方投げ）Literally, "four directions (shiho) projection (nage)". A projection or immobilization technique based on under-the-arm pivot.

**Shomen Uchi**（正面打ち）Straight strike to the head.

**Shuto**（手刀）Literally, the saber (to) hand (shu). Bare hand techniques using the hand as if it were a saber. In Aikido, we use essentially two of them: Shomen Uchi* and Yokomen Uchi. We also prefer an alternative pronunciation: "te-katana" ("te" as in "karate").

**Tai Sabaki**（体捌き）The art of moving about, both in offensive and defensive mode, and often simultaneously.

**Yokomen Uchi**（横面打ち）Strike to the side of the head.

### B    Notes on Games and Play

#### B.1  Support and Collaboration

My particular appreciation for support and collaboration, as opposed to fight and confrontation, also helped me understand my position toward team sports. For as long as I can remember, I have never been interested in team sports, never enjoyed playing them, neither watching them, expect for one: volleyball. It took me a long while to understand why. The reason is that in most team sports (rugby, soccer, handball, basketball *etc.*) there is a direct confrontation with the opponent, and in order to win, you need to think of destroying your opponent's strategy as much as developing your own. So there *is* a fight. Volleyball is different because you never really face your opponent directly or physically. Instead, because there is so little confrontation, you need to focus more on developing the best strategy within your own team. In other words, collaboration and construction (within your own team) is more important than destruction (of the other team's work).

#### B.2  Simplicity

Simplicity is also what drives my taste in games. I never liked chess, although I could enjoy a game of checkers from time to time. The rules are much simpler. In a similar vein, I never was interested in board, strategy, or role-play games (board games are more like "bored games" to me), nor in any game requiring more than a minute to understand the rules. On the rare occasions I played video games when I was younger, I mostly enjoyed FPS (First Person Shooter) games such as Doom or Quake. The rules are simple there: basically you shoot everything that moves…

#### B.3  Transgression

Even though I don't like games much, I still agree to play those from time to time, in an effort to be sociable, which by the way, is a conformant behavior. I am (not really) ashamed to confess however that when I agree to play a game, the pleasure I get from playing is largely increased when I cheat. I am also (not so much) ashamed to confess that the pleasure I get from cheating is nothing compared to the pleasure I get from confessing afterwards that I cheated…

### C    Transgression Examples

#### C.1  Lisp

The first example is in the area of syntax, which, in the case of Common Lisp, is almost entirely under the control of the programmer. The current syntactic state of Lisp is described in a so-called *readtable*. Readtables associate characters, called *macro characters*, with specific behavior implemented via hooks into the parser. For

■ **Listing 1**  A sample customization theme for Clon

```
1  :background blue
2  :face { option :foreground white
3                 :face { syntax :underline t :foreground red }
4                 :face { usage :foreground magenta }}
```

■ **Listing 2**  The almighty `loop` macro

```
1  (loop repeat 5
2        for x = 0 then y
3        for y = 1 then (+ x y)
4        collect y)
```

the familiar reader, the idea of macro characters is in fact very close to TEX's notion of *active characters*. Listing 1 shows a sample *theme* describing customizations for terminal output in a command-line options management library [99]. This is genuine Lisp code. What happens under the hood is that the curly braces are programmed to trigger a specific syntactic process: they recursively collect their contents as a parse-(sub)tree (which in Lisp is in fact just a list of lexical tokens or sub-lists thereof), and return a modified version of it representing a call to a function constructing `face` objects (`make-face`). One cannot begin to imagine the fun I had in designing this surface syntax on top of Lisp. After all, once you have introduced curly braces in the language, how more transgressive can you get, really?

The second example is in the area of macros. Before anything else, macros in Common Lisp are just Lisp functions. They have, however, a couple of specificities, one of which is that they do not evaluate their arguments right away, letting the macro code manipulate them in every imaginable way. As a consequence, it is possible to embed a completely new language within a macro call. Consider for instance the standard `loop` macro, as illustrated in listing 2, forget the outer macro call (`loop ...`) for a moment, and compare this to the code in appendix F.1. What you get is essentially an Algol-like DSL for iteration (granted, the prefix call to + in this example is reminiscent of regular Lisp).

Finally, let us mention the case of CLOS [12, 54]. We won't give a specific example here, but instead, just mention that when architected around the MOP [55, 72], CLOS turns out to be implemented in itself. For example, a CLOS class is itself an instance of some class (a *meta-class*). Similarly the whole semantics of CLOS is implemented by means of generic functions and methods specialized on standard classes. From this, it follows that by subclassing standard classes and specializing standard generic functions furthermore (or even overriding the standard methods), it is possible to effectively extend or modify the original semantics of the object layer.

The examples above can be seen merely as a collection of language tools and techniques to implement a more profound concept: that of *behavioral reflection*. Reflexivity is often confused with *introspection*, that is, the ability to examine yourself. However, there is another important aspect of reflexivity, called *intersession*: the ability to modify yourself. An even deeper distinction lies in the notions of *structural vs. behavioral* reflection [64, 81]. While structural reflection deals with providing a way to reify a program, behavioral reflection deals with accessing the language itself. Extending the Lisp syntax is a form of behavioral intercession at the parser level. Using Lisp macros is a form of structural intercession providing a *homogeneous meta-programming system* [80]: (pieces of) programs are manipulated in the language itself, contrary to C++ templates for instance, which are heterogeneous: a different language. Finally, using the CLOS MOP is a form of behavioral intercession at the object system level.

## C.2 Jazz

One particularly transgressive concept in Jazz improvisation is the idea of "playing out". At some point, an improviser uses a scale or a set of scales that do not correspond to the underlying harmonic context (for instance, playing in a different key), hence introducing dissonances that we call *frictions* in our jargon. Playing out lets you follow the song, then leave it, and get back to it again, not unlike the cat bouncing and jumping in all directions, and yet, always falling back on his feet. Some contemporary Jazz musicians like Randy Brecker are notorious for their mastery of such techniques. Perhaps the idea of playing out originated with Miles Davis, in his early electric period, where he started using chromaticisms such as minor $3^{rds}$ on top of major chords or the other way around. A very bluesy thing to do, but also very dissonant in some contexts. At the time, some of the Jazz purists mentioned earlier wanted to burn the heretic alive for even daring to try out such things.

To draw a bridge with the notion of collaboration that we developed in the Conformance section, note that playing out works even more efficiently when the rhythm section of the band accompanies the improviser in his journey abroad, by adapting in real-time the underlying harmony or rhythm to his unexpected discourse. Remember Herbie Hancock narrating how Miles Davis made his harmonic mistakes right by adapting his lead playing {13}? What happens here is exactly the opposite: whether intentional or not, a dissonant improvisation can be made right simply by adapting the underlying harmony to it. This is a game Jazz musicians love to play, and this explains why in Jazz, there is fundamentally no wrong note [10, 102]. Here again, we see how Jazz is all about adapting the musical language to your desires, more than the other way around.

## C.3 Aikido

The first example lies in kicks. To the question "do you have kicks in the discipline?", the correct answer is "we don't normally use them, but it's possible". In fact, many of the techniques to respond to hand attacks also work on kicks. What this means is that even though we don't really practice kicks, transgressing the set of codified attacks to use the foot is possible.

The way we organize practice sessions is another example of our concern to maintain a permanent state of transgression in the discipline. In Aikido, everyone practices with everyone, without any distinction of grade, skill, weight, age, or anything. On top of that, we switch partners very frequently. The whole purpose here is to never become accustomed to a partner, always ready to face the unexpected or unorthodox. In particular, and this may sound quite surprising to practitioners of other martial arts, experienced Aikidoka consider it very enriching to practice with beginners. The reason is that a beginner will often behave in very unexpected ways, even ways that are incorrect from a purely martial point of view (for example, attacking on the wrong side, completely exposing oneself). Even though the attack may be considered incorrect, we still need to be able to react to it properly. Note how reacting properly to an incorrect attack is very reminiscent of Miles Davis improvising properly on Herbie Hancock's incorrect chord {13}, or a rhythm section adjusting to the dissonance of an improvisation (appendix C.2).

One final example of a cultivated transgression in Aikido is the so-called Kaeshi Waza*. Kaeshi Waza is a transgression of the rule of collaboration and support that is a central concept in Aikido. Normally, it is considered very wrong for an experienced Aikidoka to block a beginner. Blocking doesn't help in improving one's technique and feeling the right movement physically. It only is a rather brutal way of saying "you're wrong!". So in general, an experienced Aikidoka will help beginners by making them feel the correct physical placement they should have adopted, had they mastered the technique. When experienced Aikidoka work together however, it is allowed to take the opportunity of a weakness or an opening in the partner's movement to reverse the situation and regain control. This is what is called Kaeshi Waza, and it is fun to do.

**D** **Score Example 1: Corridor**

*Excerpt from one of my most complicated scores (notation-wise).*

Corridor - page 3/4

**E** **Score Example 2: Vert de Blues**

*Full rendition of one of my simplest scores (notation-wise).*

**F** **Comparative Uniformity**

### F.1 Common Lisp Example

```
1  (let ((cmdline-options (list)))
2    (do ((cmdline-option
3           (pop (cmdline-options context))
4           (pop (cmdline-options context))))
5         ((null cmdline-option))
6      (cond ((eq (cmdline-option-option cmdline-option) option)
7             (setf (cmdline-options context)
8                   (nreconc cmdline-options (cmdline-options context)))
9             (return-from getopt
10               (values (cmdline-option-value cmdline-option)
11                       (list (cmdline-option-source cmdline-option)
12                             (cmdline-option-name cmdline-option)))))
13            (t
14             (push cmdline-option cmdline-options)))))
15    (setf (cmdline-options context) (nreverse cmdline-options)))
```

### F.2 C++ Example

```
1  template <template <class> class M, typename T, typename V>
2  struct ch_value_ <M <tag::value_<T>>, V>
3  { typedef  M<V> ret; };
4
5  template <template <class> class M, typename I, typename V>
6  struct ch_value_ <M <tag::image_<I>>, V>
7  { typedef  M <mln_ch_value(I, V)> ret; };
8
9  template <template <class, class> class M, typename T,
10   typename I, typename V>
11  struct ch_value_ <M <tag::value_<T>, tag::image_<I>>, V>
12  { typedef  mln_ch_value(I, V) ret; };
13
14  template <template <class, class> class M, typename P,
15   typename T, typename V>
16  struct ch_value_ <M <tag::psite_<P>, tag::value_<T>>, V>
17  { typedef  M<P, V> ret; };
```

**About the author**

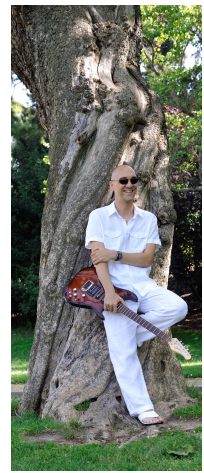

**Didier Verna** Half of the time, Didier Verna works as an associate professor in computer science. He's the president of the European Lisp Symposium steering committee and organizes the event every year. Contact him at didier@lrde.epita.fr.

The other half of the time, he works as a Jazz guitarist and composer. He has released two albums of original compositions with his band: the @-quartet. Contact him at didier@didierverna.com.

The third (and slightly shorter) half of the time, he also is an Aikido practitioner and gives coaching sessions on the theme "Aikido and Conflict Management" for large companies. He holds a federal Aikido teaching certificate. Contact him at didousan@didierverna.org.